\documentclass{article}
\usepackage{spconf,amsmath,graphicx}
\usepackage{multirow}
\usepackage{multicol}
\usepackage{array}
\usepackage{booktabs}
\usepackage[colorlinks,
            linkcolor=black,
            anchorcolor=black,
            citecolor=black]{hyperref}
\usepackage{fancyhdr}

\title{The Multimodal Information based Speech Processing (Misp) 2023 Challenge: Audio-Visual Target Speaker Extraction}
\name{
\begin{tabular}{c}
Shilong Wu\textsuperscript{1}, 
Chenxi Wang\textsuperscript{1},
Hang Chen\textsuperscript{1},
Yusheng Dai\textsuperscript{1},
Chenyue Zhang\textsuperscript{1}, \\
Ruoyu Wang\textsuperscript{1}, 
Hongbo Lan\textsuperscript{1},
Jun Du\textsuperscript{1}, 
Chin-Hui Lee\textsuperscript{2}, 
\textit{Jingdong Chen\textsuperscript{3}},  
\textit{Shinji Watanabe\textsuperscript{4}},\\  
\textit{Sabato Marco Siniscalchi\textsuperscript{2,5}}, 
\textit{Odette Scharenborg\textsuperscript{6}}, 
\textit{Zhong-Qiu Wang\textsuperscript{4}}, 
\textit{Jia Pan\textsuperscript{7}}, 
\textit{Jianqing Gao\textsuperscript{7}}
\end{tabular}
} 
\address{
 \textsuperscript{1} University of Science and Technology of China, China  \textsuperscript{2} Georgia Institute of Technology, USA\\
 \textsuperscript{3} Northwestern Polytechnical University, China 
 \textsuperscript{4} Carnegie Mellon University, USA \\
 \textsuperscript{5} Kore University of Enna, Italy 
 \textsuperscript{6} Delft University of Technology, The Netherlands
 \textsuperscript{7} iFlytek, China\\ 
}
\begin{document}
\ninept
\maketitle

%\thispagestyle{fancy}

%\lfoot{© 20XX IEEE. Personal use of this material is permitted. Permission from IEEE must be obtained for all other uses, in any current or future media, including reprinting/republishing this material for advertising or promotional purposes, creating new collective works, for resale or redistribution to servers or lists, or reuse of any copyrighted component of this work in other works.}
%\cfoot{}
%\renewcommand{\headrulewidth}{4mm}

%
\begin{abstract}
Previous Multimodal Information based Speech Processing (MISP) challenges mainly focused on audio-visual speech recognition (AVSR) with commendable success. However, the most advanced back-end recognition systems often hit performance limits due to the complex acoustic environments. This has prompted a shift in focus towards the Audio-Visual Target Speaker Extraction (AVTSE) task for the MISP 2023 challenge in ICASSP 2024 Signal Processing Grand Challenges. Unlike existing audio-visual speech enhancement challenges primarily focused on simulation data, the MISP 2023 challenge uniquely explores how front-end speech processing, combined with visual clues, impacts back-end tasks in real-world scenarios. This pioneering effort aims to set the first benchmark for the AVTSE task, offering fresh insights into enhancing the accuracy of back-end speech recognition systems through AVTSE in challenging and real acoustic environments. This paper delivers a thorough overview of the task setting, dataset, and baseline system of the MISP 2023 challenge. It also includes an in-depth analysis of the challenges participants may encounter. The experimental results highlight the demanding nature of this task, and we look forward to the innovative solutions participants will bring forward.

\end{abstract}

\section{Introduction}
\label{sec:intro}
In real-world scenarios, the complex and adverse acoustic environment is one of the main challenges of automatic speech recognition (ASR) and other back-end tasks. The strong noise, reverberation, and multi-speaker interference result in a serious impact on the system performance. 
Effective front-end speech processing technologies, like speech enhancement and speech separation \cite{front1, front2}, have been proven to play a significant role in improving speech quality,  thereby enhancing the performance of back-end systems. Recently, research on the cocktail-party problem \cite{cocktail} has shown that people can naturally track the speech of the target speaker from the interference of multiple speakers' conversations and background noise. Inspired by this, researchers have begun to focus on target speaker extraction (TSE), which estimates the speech of the target speaker within a mixed audio stream, leveraging a diverse array of clues including auditory, spatial, visual, and other \cite{TSE}.

Research on TSE systems has mainly focused on audio-only TSE (AOTSE) \cite{AOTSE1, AOTSE2}, which utilizes the pre-registered speech from speakers as clues. This method is marked by its inherent simplicity, as it circumvents the need for supplementary equipment. Nonetheless, the development of AOTSE is constrained by several factors, including the challenges associated with acquiring pre-registered audio in real-world scenarios, the potential similarities of acoustic features among multiple speakers, and the presence of significant noise interference \cite{AOTSE3}. Recently, research \cite{visual} in neuroscience suggests that the visual modality, including facial and lip movements, can significantly influence humans' auditory attention, enhancing speech perception by providing additional information about the speaker, especially in noisy environments \cite{mease}. In real-world scenarios, acquiring visual clues is common, which can effectively overcome the challenges encountered by AOTSE. As a result, an increasing number of researches focus on audio-visual TSE (AVTSE) \cite{AVTSE1, AVTSE2}. Unfortunately, no publicly available benchmark currently exists for AVTSE. To fill this gap, the Multi-modal Information based Speech Processing (MISP) 2023 challenge focuses on the AVTSE task.

\begin{figure*}[htbp]
\centering
\includegraphics[width=0.85\textwidth]{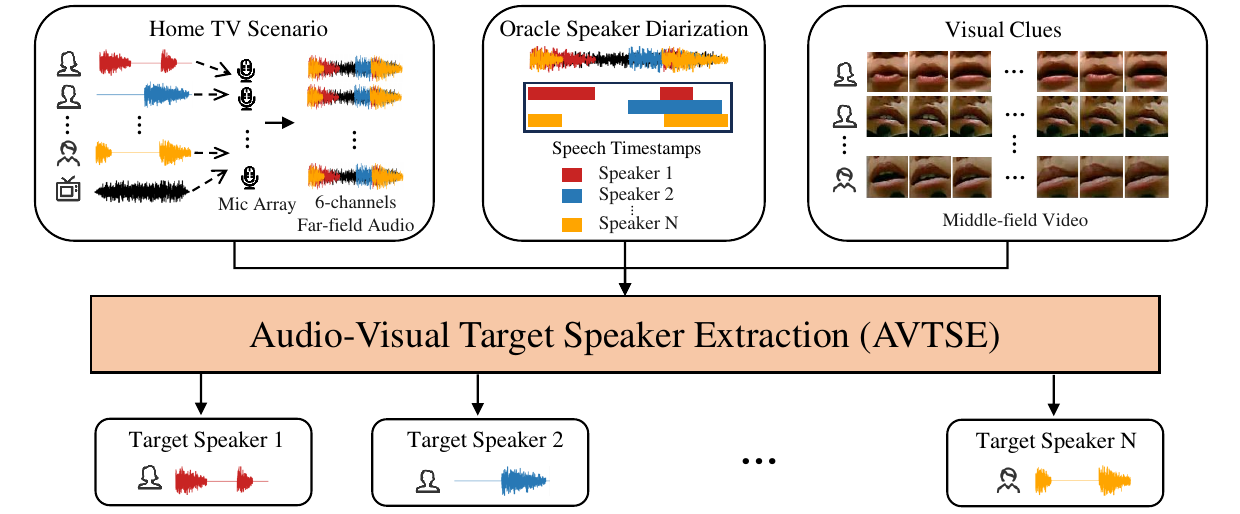}
\vspace{-0.3cm}
\caption{Overview of the AVTSE task in MISP 2023 challenge.}
\label{fig: overview}
\vspace{-0.6cm}
\end{figure*}

Expanding the scope to front-end speech processing technology, there are already several relevant challenges within the field of speech enhancement. These include the audio-only speech enhancement (AOSE) challenges like the Deep Noise Suppression (DNS) Challenge \cite{DNS}, the Clarity Challenge \cite{cla}, and multimodal challenges such as the COG-MHEAR Audio-Visual Speech Enhancement Challenge \cite{cog}. Nevertheless, there are two main problems with current challenges. Firstly, the evaluation data is either the simulation data obtained by adding a single type of noise or interference speech to clean speech, or it is recorded in a real scene but speakers just read specific sentences or word arrangements. However, in real-life scenarios, people's conversations typically do not have a specific topic, and they encounter complex acoustic environments with multiple types of noise, reverberation, and interference from other speakers, which can lead to a mismatch between simulation and reality. Secondly, these challenges often utilize metrics such as the Deep Noise Suppression Mean Opinion Score (DNSMOS) \cite{DNSMOS}, the Short-Time Objective Intelligibility (STOI) \cite{STQI}, and the Perceptual Evaluation of Speech Quality (PESQ) \cite{PESQ} to evaluate speech quality, or invite staff to score based on their actual listening experience. Research has shown that enhanced speech may experience distortion in terms of comprehensibility, which will worsen the performance of back-end ASR systems \cite{se2asr}. Therefore, evaluating its impact on the performance of back-end systems is also extremely important.

%the Hurricane Challenge \cite{hur}

In the previous MISP challenges \cite{MISP2021, MISP2022}, we released a large distant multi-microphone conversational Chinese audio-visual corpus that focuses on real home-TV scenarios, where 2-6 speakers freely converse without specific topics. Building upon the above analysis, the MISP 2023 challenge aims to improve the accuracy of the back-end ASR system through the AVTSE system in real-world scenarios using the MISP corpus. Specifically, we will use a pre-trained ASR model to decode the speech output from the AVTSE system, with character error rate (CER) as the evaluation metric. Through this challenge, we hope to promote researchers' attention to the AVTSE system, and provide new ideas for the front-end technology application in real scenarios and joint optimization with back-end systems. In this paper, we will introduce the task setting, dataset, and baseline system of the MISP 2023 challenge and conduct an in-depth analysis of the potential difficulties faced by participants. More details can be found on the website\footnote{https://mispchallenge.github.io/mispchallenge2023}.

\vspace{-0.2cm}
\section{Dataset and Task Setting}
\label{sec:task}
\vspace{-0.15cm}
\subsection{Dataset Scenarios and Composition}
The MISP corpus \cite{MISP2021} focuses on real home-TV scenarios: 2-6 people communicating with each other with TV noise and reverberation in the background. In this scenario, speakers engage in spontaneous conversations without specific topics, posing a challenge due to the significant speech overlap and diversity. Furthermore, in some sessions, strong background noise from television is present, where television programs such as dramas, news, music, and interviews may be playing, further exacerbating the complexity, especially for front-end systems.

We use the training set of AVSR corpus of MISP 2021 challenge \cite{MISP2021-2}, with a duration of 106.09 hours, including 21 rooms and 200 speakers. In the training set, the data includes near/middle/far-field audio and middle/far-field videos, allowing participants to choose freely. Additionally, we use the development set released in the MISP 2022 challenge \cite{MISP2022}, with a duration of 2.51 hours. For the evaluation set, in addition to the MISP 2022 evaluation set, we will also add some new sessions focusing on female dialogue scenarios. Our research has revealed that distinguishing voices in female dialogue scenarios is more challenging, thus increasing the difficulty level for participants. For development and evaluation sets, participants can only use far-field audio and middle-field video.

\thispagestyle{empty}

%Due to the differences in crystal clocks of recording devices, there may be misalignment between near-field and far-field audio. Therefore, for the AVTSE task, we recommend that participants use the near-field audio to simulate the far-field audio and use the near-field audio as labels to train the model, or directly use far-field data for joint training with the back-end AVSR model.
\vspace{-0.2cm}
\subsection{Task Overview and Evaluation}

The MISP 2023 challenge focuses on the AVTSE task, as shown in \autoref{fig: overview}. Participants must utilize multi-channel far-field audio along with the target speaker's middle-field video to extract the target speaker's speech from audio recordings containing overlapping voices of multiple speakers and background noise. In one session, each speaker is sequentially treated as the target speaker. In addition, we will also provide the oracle diarization results because this is the first edition of the AVTSE challenge, and not providing them would bring greater difficulties to participants. This year, we will open two rankings based on whether external data, apart from the MISP dataset, has been utilized. However, it is worth noting that we will only submit the ranking which is without using them to ICASSP SPGC. This is because we encourage technological innovation rather than relying on large amounts of data.

The objective of the MISP 2023 challenge is to enhance the front-end AVTSE system to extract clearer speech of the target speaker, thereby improving the accuracy of speech recognition while keeping the back-end ASR system unaltered. We will provide a pre-trained ASR model \cite{AVSR}, including both model parameters and code. To better investigate the role of the AVTSE system, we only use the audio-only ASR part of that paper. Participants can test the extracted speech in the development set to make adjustments to the AVTSE model. For ranking, we use the character error rate (CER) as the evaluation metric:
\vspace{-0.1cm}
\begin{equation}
    \textrm{CER} = {(S+D+I)}/{N}\times 100\%
\vspace{-0.1cm}
\end{equation}
where, $S$, $D$, and $I$ represent the number of substitutions, deletions, and insertions. $N$ is the number of characters in ground truth. The lower the CER, the higher the ranking.

During the evaluation stage, participants must submit the extracted speech, which we will decode and use to calculate CER. Therefore, participants do not need to modify the ASR model. However, they can still conduct joint training on the front-end and back-end systems with fixed back-end parameters. This is also an avenue we encourage participants to explore. Furthermore,  we will calculate the DNSMOS P.835 \cite{DNSMOS} as a reference to explore the relationship between speech auditory quality and back-end tasks.

\vspace{-0.2cm}
\section{Baseline System}
\label{sec:baseline}

\begin{figure}[t]
\centering
\includegraphics[width=\linewidth]{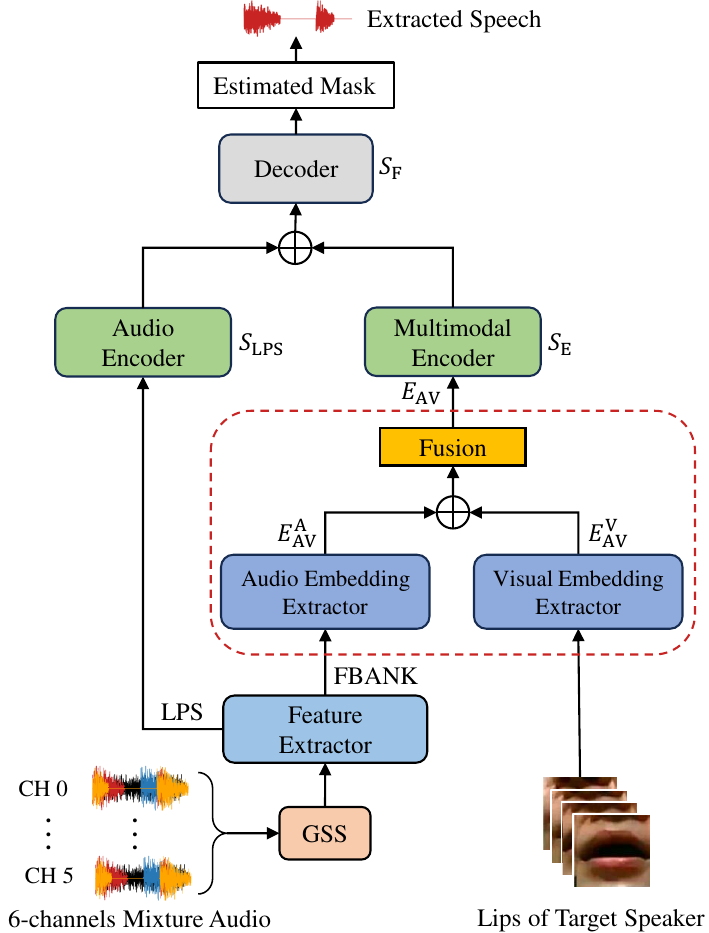}
\vspace{-0.65cm}
\caption{Diagram of the baseline system.}
\label{baseline}
\vspace{-0.65cm}
\end{figure}

\vspace{-0.2cm}
\subsection{Data Preparation}
\vspace{-0.1cm}
To achieve complete alignment between speech signals, our initial step involves utilizing near-field data to simulate far-field conditions as the training data. Owing to potential noise interference or unclear speech in certain sections of the raw near-field audio data, it is imperative to conduct data cleaning on the raw training set. Firstly, we segment the near-field speech based on timestamps, ensuring that each segment contains the speech of a single speaker. Then we employ the DNSMOS P.835 to identify segments with high speech quality. We meticulously screen the training set, retaining only those near-field speech segments with high overall quality (OVRL) scores. This rigorous selection process results in the inclusion of a total of 22.13 hours of high-quality near-field speech segments. For model training, we randomly combine near-field speech and add noise with different signal-to-noise ratios (SNR) from -10dB to +20dB and reverberation related to room parameters to simulate 6-channels far-field audio. To minimize the mismatch with real data as comprehensively as possible, the noise we utilize is extracted from the raw training set, containing a large amount of interference from the speaker's voice, as well as environmental and TV noise.

\vspace{-0.2cm}
\subsection{System Architecture}
\vspace{-0.1cm}
As shown in \autoref{baseline}, the baseline system is mainly based on our recent work - the multimodal embedding aware speech enhancement (MEASE) \cite{mease} model, which has achieved "SOTA" in the field of audio-visual speech enhancement (AVSE). Building upon this foundation, we leveraged the oracle diarization results to conduct guided source separation (GSS) \cite{GSS} on 6-channels mixture audio to initially mitigate the impact of the overlapping speech. Then we use the MEASE model to further extract the speech of the target speaker. The MEASE model comprises a multimodal embedding extractor (in red dashed box) and an embedding-aware enhancement network. We first extract FBANK features and noisy log-power spectra (LPS) features from the audio output of GSS. Subsequently, we use the pre-trained embedding extractor to obtain the deep embeddings from both the FBANK ($A_{\rm FBANK}$) and lip frames ($V$) of the target speaker. Both audio and visual embedding extractors consist of a spatiotemporal convolution followed by an 18-layer ResNet, but the structure is slightly different and can be represented as follows:

\vspace{-0.3cm}
\begin{equation}
\begin{split}
\begin{aligned}
&E_{\rm AV}^{\rm A} = \rm ResNet18_{1D}(BN(ReLU(Conv_{1D}(\textit{A}_{FBANK}))))\\
&E_{\rm AV}^{\rm V} = \rm ResNet18_{2D}(MP_{3D}(BN(ReLU(Conv_{3D}(\textit{V})))))
\end{aligned}
\end{split}
\end{equation}
where $\rm ReLu(\cdot)$, $\rm BN(\cdot)$ and $\rm MP_{3D}(\cdot)$ represents the ReLU activation, the batch normalization, and the spatiotemporal max-pooling layer, respectively. Then, we fused the audio and visual embeddings using 2-layers of BiGUR. For encoders and decoder, we use the different numbers of 1D-ConvBlocks, where, $N_{\rm LPS}$=5, $N_{\rm E}$=10, and $N_{\rm F}$=15. At the end of the decoder, the hidden representation is activated by a sigmoid activation to generate a magnitude mask. We utilize the ideal ratio mask (IRM) \cite{irm} as the learning target and the mean square error (MSE) $\mathcal{L}_{\rm{MSE}}$ between target IRM and estimated mask as the loss function. 
\thispagestyle{empty}

\begin{figure}[t]
\centering
\includegraphics[width=\linewidth]{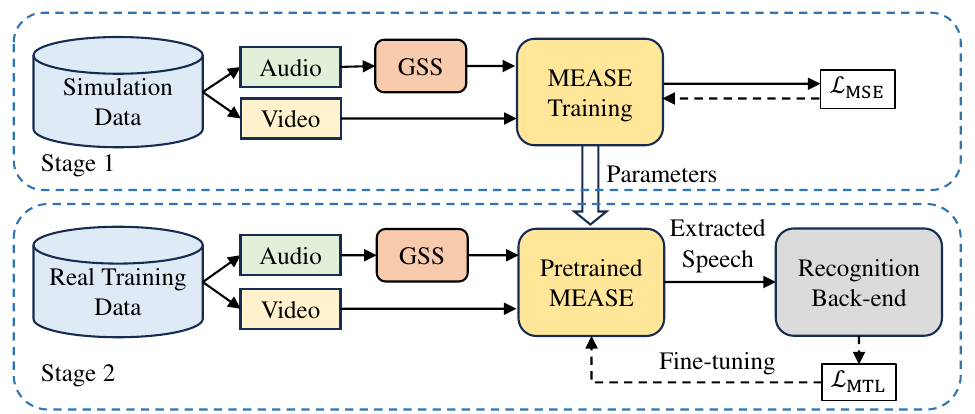}
\vspace{-0.6cm}
\caption{The two-stage training process of the baseline system.}
\label{twostage}
\vspace{-0.6cm}
\end{figure}

\vspace{-0.2cm}
\subsection{Training Process}
\vspace{-0.1cm}
As shown in \autoref{twostage}, the training process of the baseline system encompasses two stages. Firstly, we trained the MEASE model using the simulated data with $\mathcal{L}_{\rm{MSE}}$ as the loss function. However, this training approach inevitably leads to some degree of distortion in the extracted speech since it does not consider the back-end recognition task, thereby impacting the accuracy of the recognition system. Consequently, in the second stage, we fine-tuned the pre-trained MEASE model using the recognition back-end. Furthermore, to mitigate the issue of mismatch between simulated data and real-world scenarios, we utilized the real far-field data from the training set in the second stage. Because we require that the parameters of the back-end system cannot be changed, we freeze the recognition model and only use the loss returned by it. In our provided ASR back-end \cite{AVSR}, the loss function employed is denoted as $\mathcal{L}_{\rm{MTL}}$:
\vspace{-0.05cm}
\begin{equation}
    \mathcal{L}_{\rm {MTL}} = \lambda {\rm log} P_{\rm {ctc}}(Y|X)+(1-\lambda){\rm log} P_{\rm {att}}(Y|X)
    \vspace{-0.05cm}
\end{equation}
where $X$ and $Y$ denote the encoder output and the target sequences, respectively. $\lambda$ is the weight factor between the CTC loss and the attention cross entropy (CE) loss \cite{ctc/att}. Here we set $\lambda = 0.3$.

%\begin{table*}[!t] 
%\centering
%\renewcommand\arraystretch{1.2} 
%\setlength\tabcolsep{17pt}
%\caption{Detailed CER (\%) and DNSMOS results on development set}
%\vspace{3pt}
%\label{tab:baseline}
%\begin{tabular}{lccccccc}
%\toprule
%\multirow{2}{*}{Method} & \multicolumn{4}{c}{Recognition Backend}                    & \multicolumn{3}{c}%{DNSMOS} \\ \cmidrule(r){2-5} \cmidrule(r){6-8} 
%                        & S    & D   & {I}   & CER  & SIG     & BAK     & OVR    \\ 
%\cmidrule(r){1-1} \cmidrule(r){2-5} \cmidrule(r){6-8}  
%Near-field Audio        & 10.8 & 1.9 & {1.1} & 13.8 & 2.98    & 3.11    & 2.43   \\ 
%Beamforming             & 23.1 & 4.7 & {3.0} & 30.8 & 1.45    & 1.35    & 1.20   \\ 
%GSS                     & 15.8 & 3.1 & {1.9} & 20.7 & 1.78    & 1.80    & 1.35   \\ 
%GSS+AEASE               & 17.4 & 3.7 & {2.1} & 23.2 & 1.97    & 2.08    & 1.43   \\ 
%GSS+MEASE               & 16.2 & 3.3 & {1.9} & 21.4 & 2.01    & 2.14    & 1.46   \\ 
%GSS+MEASE+Finetune      & 16.0 & 3.2 & {1.8} & 21.0 & 1.83    & 1.86    & 1.38   \\ 
%\bottomrule
%\end{tabular}
%\vspace{-0.4cm}
%\end{table*}

\begin{table*}[!t] 
\centering
\renewcommand\arraystretch{1.1} 
\setlength\tabcolsep{18pt}
\caption{Detailed CER (\%) and DNSMOS results of different front-end systems on the development set.}
\vspace{3pt}
\label{tab:baseline}
\begin{tabular}{lccccccc}
\toprule
\multirow{2}{*}{System} & \multicolumn{4}{c}{Recognition Back-end}                    & \multicolumn{3}{c}{DNSMOS} \\ \cmidrule(r){2-5} \cmidrule(r){6-8} 
                        & S    & D   & {I}   & CER  & SIG     & BAK     & OVR    \\ 
\cmidrule(r){1-1} \cmidrule(r){2-5} \cmidrule(r){6-8}  

Beamforming             & 31.0 & 7.2 & 4.8 & 43.0 & 1.45    & 1.35    & 1.20   \\ 
GSS                     & 20.0 & 4.2 & 2.2 & 26.4 & 1.78    & 1.80    & 1.35   \\ 
GSS+AEASE               & 21.5 & 4.9 & 2.2 & 28.6 & 1.97    & 2.08    & 1.43   \\ 
GSS+MEASE               & 20.4 & 4.4 & 2.2 & 27.0 & 2.01    & 2.14    & 1.46   \\ 
GSS+MEASE+Finetune      & 19.8 & 4.7 & 1.8 & 26.3 & 2.03    & 2.27    & 1.50   \\ 
\bottomrule
\end{tabular}
\vspace{-0.5cm}
\end{table*}

\vspace{-0.1cm}
\section{RESULTS AND ANALYSIS}
\label{sec:result}
\vspace{-0.1cm}
\subsection{Baseline Results}
\vspace{-0.05cm}
\autoref{tab:baseline} shows the results of different front-end systems in the speech recognition evaluation metric and DNSMOS P.835. Among these systems, AEASE is a simplified version of MEASE, as it does not utilize the visual modality. The results of "GSS+MEASE+Finetune" serve as our final baseline results. First, the comparison of the results of Beamforming \cite{beamforming} and GSS proves the importance of separating the speaker's speech from the mixed audio in challenging acoustic scenes. 
Incorporating the AEASE model on top of GSS, we observed an improvement in the DNSMOS, which reflects better realistic auditory quality. However, in terms of ASR metrics, it performed worse than GSS. This observation underscores that while AEASE can filter out more noise, it also introduces speech distortion simultaneously. Replacing the AEASE model with the MEASE model, which incorporates the visual modality of the target speaker, leads to a notable enhancement in the DNSMOS. Additionally, the introduction of the visual modality can enable the model to extract more of the target speaker's speech, consequently yielding advancements in speech recognition results. Nevertheless, there remains a gap when compared to GSS. In order to balance back-end performance and noise suppression, we use the back-end recognition system to fine-tune the MEASE model. This enables the model to process signals in a targeted manner to improve the performance of the back-end system. The results show that the model can better preserve the target speaker's speech and remove more irrelevant noise, while the speech recognition results are also improved.

\begin{figure}[!t]
\centering
\includegraphics[width=\linewidth]{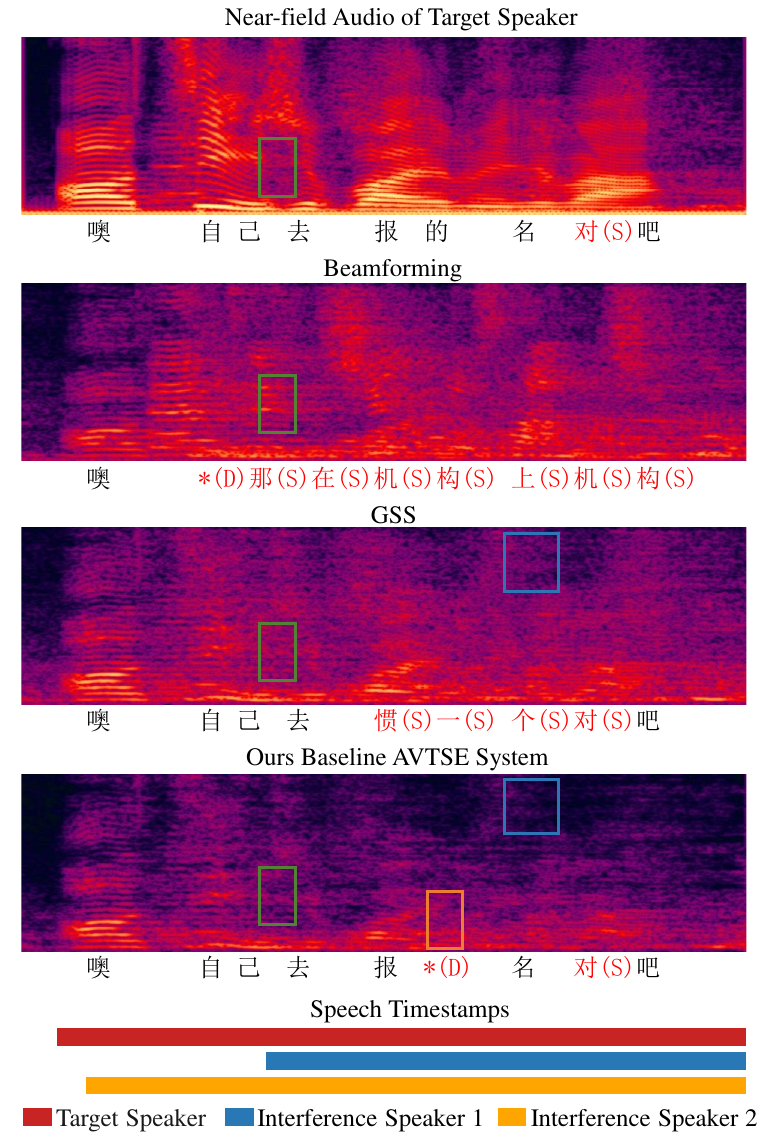}
\vspace{-0.6cm}
\caption{Spectrograms and recognition results of an example.}
\label{example}
\vspace{-0.6cm}
\end{figure}

However, although our baseline system surpasses GSS in terms of DNSMOS, its speech recognition performance is only comparable to that of GSS. Due to noise suppression, it is inevitable to introduce more deletion errors. This is sufficient to demonstrate that in real scenarios, our task is very challenging. Therefore, how to improve the AVTSE system to enhance the accuracy of the backend recognition system requires more in-depth research by participants.

\vspace{-0.2cm}
\subsection{Difficulty Analysis}
\vspace{-0.05cm}

\autoref{example} is an example of a mixture with multi-speaker interference and strong TV background noise. According to the speech timestamps, it can be seen that this is a complex acoustic scene with 2 to 3 speakers speaking simultaneously. We compared the spectrograms and the recognition results obtained from different systems. Firstly, we directly perform ASR decoding on near-field audio and obtain the result that is likely to be at the theoretical limit. Then, we compared beamforming, GSS, and baseline AVTSE system for far-field audio. According to beamforming's spectrogram, it is evident that this is a highly noisy environment so the backend system cannot distinguish the target speaker's speech, resulting in a very poor recognition result. In contrast, GSS can separate the speech of the target speaker but still contains a significant amount of noise. Our baseline AVTSE system further reduced noise, resulting in the best recognition results on far-field audio. However, there is still a significant quality gap between the extracted speech and near-field audio.

\thispagestyle{empty}

As shown in the green boxes in \autoref{example}, there is a notable presence of the interference speakers' speech in the far-field audio, affecting the clarity of the speaker's speech. However, due to the influence of the visual modality, AVTSE has the capability to filter out this interference. As demonstrated in the blue boxes, the certain background noise that remains in GSS output will largely be eliminated by AVTSE, consequently correcting the back-end recognition results. Nevertheless, as shown in the yellow box, our baseline system inevitably filters out some vocal components while suppressing noise, resulting in deletion errors. Therefore, there is still significant potential for improvement within the AVTSE system. Extracting the target speaker's speech while suppressing noise is challenging.

\vspace{-0.05cm}
\section{CONCLUSIONS}
\label{sec:conclusions}
\vspace{-0.1cm}
In this paper, we provide a detailed description of the dataset, task setting, and baseline system for the MISP 2023 challenge, which is the first benchmark of the AVTSE task. We also conducted a deep analysis of the baseline experimental results, highlighting that the AVTSE task continues to hold significant research potential in real-world scenarios. In the future, we plan to explore the solutions for AVTSE systems on long recordings and incorporate the subjective listening test to further research the relationship between the real speech auditory quality and the performance of back-end tasks.

%We anticipate that the MISP 2023 challenge will contribute to the advancement of the AVTSE field.

% -------------------------------------------------------------------------

\clearpage
%\vfill
%\pagebreak
% -------------------------------------------------------------------------
\thispagestyle{empty}
\bibliographystyle{IEEEbib}
\bibliography{Main}

\end{document}